\documentclass[epj,referee]{svjour}
\usepackage{graphicx}
\usepackage{amsmath}
\usepackage{amssymb}
\usepackage{color}

\newcommand{\ii}{\text{i}}
\newcommand{\ee}{\text{e}}

\newcommand{\of}[1]{\left ({#1}\right)}
\newcommand{\dd}{\text{d}}

\newcommand{\fracp}[2]{\frac{\partial {#1}}{\partial {#2}}}
\newcommand{\eigengamma}{\gamma}

\begin{document}
\title{How Coupling Determines the Entrainment of Circadian Clocks}
\author{G. Bordyugov$^\dagger$ \and A.E. Granada$^\dagger$ \and H. Herzel
\thanks{We thank Werner Ebeling and Vadim Anishchenko for their guidance through the world of coupled oscillators.}}

\institute{
$^\dagger$ Both authors contributed equally\\
Institute for Theoretical Biology, Humboldt University, Invalidenstra\ss e 43, D-10115 Berlin, Germany}

\date{\today}

\abstract{
Autonomous  circadian clocks drive daily rhythms in physiology and behaviour. A network of coupled neurons, the suprachiasmatic nucleus (SCN), serves as a robust self-sustained circadian pacemaker.
Synchronization of this timer to the environmental light-dark cycle is crucial for an organism's fitness.
In a recent theoretical and experimental study it was shown that coupling governs the entrainment range of circadian clocks.
We apply the theory of coupled oscillators to analyse how diffusive and mean-field coupling affects the entrainment range of interacting cells.
Mean-field coupling leads to amplitude expansion of weak oscillators and, as a result, reduces the entrainment range.
We also show that coupling determines the rigidity of the synchronized SCN network, i.e. the relaxation rates upon perturbation. 
Our simulations and analytical calculations using generic oscillator models help to elucidate how coupling determines the entrainment of the SCN.
Our theoretical framework helps to interpret experimental data.
}

\maketitle

\section{Introduction}
\label{Section::Introduction}
Circadian clocks are endogenous oscillators driving daily rhythms in physiology and behaviour.
In mammals, these rhythms are centrally controlled by a tiny neuronal nucleus located in the hypothalamus, the suprachiasmatic nucleus (SCN). 
SCN cells synchronize to each other and generate a robust $\sim$ 24 h self-sustained oscillation that drives locomotor and hormonal daily rhythms even in the absence of external forcing. 
The environmental day-night cycles periodically modulate these oscillations \cite{Roenneberg2003}.
Thus the circadian system can be regarded as a network of driven coupled oscillators.
Many details of the entrainment (synchronization to external environment) and intercellular synchronization processes are not well understood.
In particular, the coupling mechanisms between neurons are debated and the coupling strengths are unknown.
Here we apply the established theory of coupled  oscillators \cite{Huygens1673,Kuramoto2003,Vadim2007,Alexander2009} to study the synchronization and entrainment properties of the mammalian circadian clock.
Our goal is a better understanding how the interplay between cell-to-cell synchronization affects the entrainment properties of the whole SCN network.

This paper is organized as follows: first we summarize the major relevant physiological findings on the entrainment properties of the mammalian SCN and peripheral tissues.
In order to capture the most fundamental oscillatory features of the circadian clock, we compare those experimental findings with modeling results.
We study for this purpose rather generic coupled oscillators.
We conclude that observed differences in the entrainment range of SCN and peripheral tissues can be assigned to coupling-induced amplitude and rigidity changes.
We discuss implications for the interpretation of experimental results.

\subsection{The SCN \--- a network of coupled of oscillators}
Oscillations in the SCN emerge at the individual cell level and are generated by intracellular genetic feedback loops \cite{Ueda2010}.
Experiments with dispersed SCN neurons revealed that individual cells display oscillations with periods ranging from 20 to 28 hours \cite{Welsh1995}. 
Coupling between individual SCN neurons confers a precise and robust synchronized rhythm \cite{Herzog2004,Liu2007a}.
Diverse intercellular coupling mechanisms are believed to be responsible for the SCN synchronization: synaptic connections, gap junctions and secreted neuropeptides \cite{Aton2005}.
Quantitative details of the coupling mechanisms are still unknown  \cite{Welsh2010}.
In this work we study two basic types of coupling mechanisms: diffusive and mean-field coupling.
\textit{Diffusive coupling} is generated by a difference in state between the given cell and its neighbourhood, hence its name, whereas in \textit{mean-field coupling} all cells equally contribute to the common mean-field, which acts back on each of them (see Section~\ref{Subsection::MeanField}).

\subsection{Entrainment of the circadian clock}
The circadian clock has evolved to synchronize an organism to periodically recurring environmental conditions, such as light-dark or temperature cycles.
Robust entrainment to external periodic signals is essential for a precise timing of behaviour and metabolism.
External stimuli such as light or neuropeptide pulses can  shift the phase of the SCN clock by a few hours \cite{Comas2006,Piggins1995}.
Periodically applied external stimuli can entrain the SCN to a diverse range of periods typically within a range of $24\pm 2$ h \cite{Pittendrigh1976,Aschoff1978,Vilaplana1997}.
In addition to the SCN, almost every cell in the human body shows circadian oscillations, such oscillators are known as peripheral circadian oscillators \cite{Liu2007a,Yagita2000}.

In a recent series of experiments, temperature cycles within the physiological range were applied to organotypic SCN and lung slices and their entrainment ranges were characterized \cite{Buhr2010,Abraham2010}.
Even though the molecular mechanism driving the oscillations at single cell level in peripheral and SCN tissues are quite similar \cite{Liu2007a,Yagita2000}, the lung tissue was able to entrain to a much wider range  \cite{Abraham2010}.
These different entrainment ranges reflect presumably differences in the intercellular coupling in the SCN and peripheral tissues.
Puzzled by these differences between the SCN and peripheral tissue clock properties, we analyze systematically how oscillator properties and coupling between the oscillators affect the entrainment range.

\section{Results}
In the current paper, we make use of both direct numerical simulations of the system on hand and of numerical bifurcation analysis.
Please see Appendix~\ref{Section::Methods} for the details of the numerical methods.
\label{Section::Results}
\subsection{What oscillator properties govern the entrainment range?}
\label{Subsection::Range}
We follow the tradition of Winfree \cite{Winfree1980}, Kronauer \cite{Kronauer1982}, and Glass \& Mackey \cite{Glass1988}
and study generic amplitude-phase oscillators.
More detailed biochemical models \cite{Leloup2003,Forger2003,Becker-Weimann2004} can be approximated by amplitude-phase models \cite{Granada2009b}.
Pure phase models \cite{Kuramoto2003} might be not sufficient since it has been shown experimentally that the amplitude in the clock oscillations is variable \cite{Yamaguchi2003,Westermark2009} and affects entrainment behaviour \cite{Pittendrigh1991,Vitaterna2006,Brown2008,vanderLeest2009}.

\begin{figure}
\includegraphics[width=\linewidth]{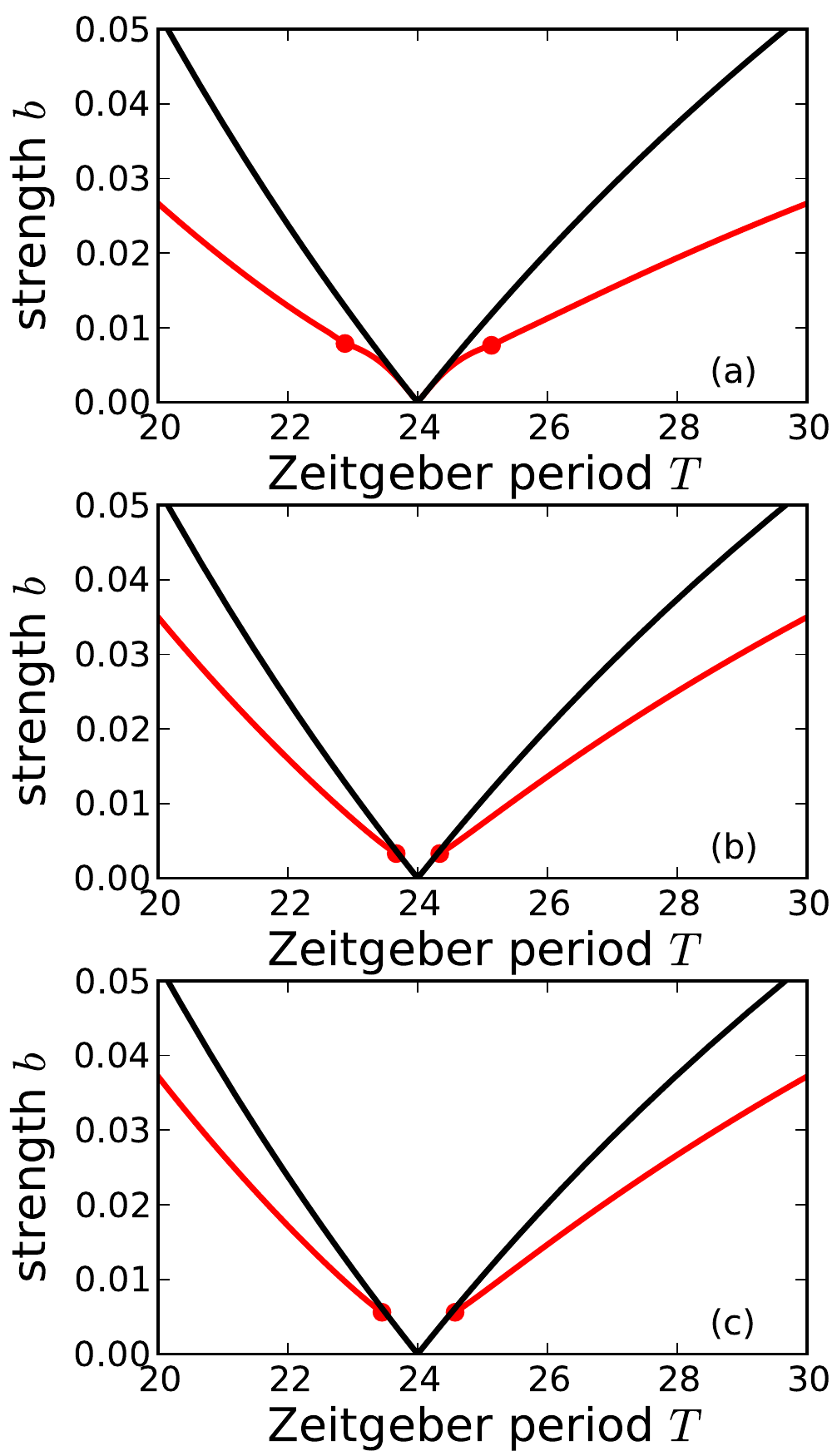}
\caption{
Entrainment range for three different single oscillator models.
In each frame, the borders of the entrainment range for rigid ($\eigengamma = 1.0$, black lines) and weak ($\eigengamma = 10^{-2}$, red lines) oscillators with $A_0 = 1.0$ and $\tau = 24$ are shown.
(a) Oscillator with linear radial dynamics $f_l$,
(b) the Poincar\'e oscillator, and (c) the Hopf oscillator.
In all three frames (a), (b), and (c), the black lines occlude the red ones in the region with small $b$.
Below red dots, red lines denote saddle-node bifurcation of limit cycles.
Above red dots, red lines denote Neimark-Sacker bifurcation of limit cycles.
Black lines always denote saddle-node bifurcation of limit cycles.
}
\label{fig::fig01}
\end{figure}

A simple amplitude-phase oscillator can be generically written as
\begin{equation}
\label{eq::general}
\begin{split}
\dot r &= - \eigengamma \, r \, f\of{r},\\
\dot \varphi &= \frac{2\pi}{\tau}.
\end{split}
\end{equation}
Here, the parameter $\tau$ denotes the intrinsic period of the oscillator.
The function $f\of{r}$ determines the particular type of oscillator.
In this paper we consider three choices of $f\of{r}$:
\begin{equation}
\label{eq::threechoices}
\begin{split}
f\of{r} &= f_h\of{r} := r^2 - A_0,\\
f\of{r} &= f_p\of{r} := r - A_0,\\
f\of{r} &= f_l\of{r} := 1 - A_0/r.
\end{split}
\end{equation}
The nonlinearity $f_h$ corresponds to a Hopf oscillator (which is sometimes called modified van der Pol oscillator, see \cite{Ebeling1986}), $f_p$ corresponds to a Poincar\'e oscillator \cite{Glass1988}, and $f_l$ represents a linear dynamics in the radial variable.
In all three cases, $A_0$ determines the size of the limit cycle.
Parameter $\eigengamma$
determines the relaxation rate towards the limit cycle with $r = A_0$.
Equation~(\ref{eq::general}) can be written in a more compact complex form as
\begin{equation}
\label{eq::complex}
\begin{split}
\dot z = \of{\ii\omega - \eigengamma f\of{r}}z
\end{split}
\end{equation}
for complex variable $z = r \ee^{\ii\varphi}$ with amplitude $r$ and phase $\varphi$.
Here, the frequency of the oscillator is given by $\omega = 2\pi/\tau$.
In this subsection, the models refer to the SCN rhythm as a whole. Coupling of single cell oscillators will be discussed below.

In the context of genetic circadian oscillators, we study mainly so-called weak oscillators, characterized by small positive $\eigengamma$, see a detailed quantification of clock oscillations in \cite{Westermark2009}.
As a consequence of small $\gamma$, the amplitude of weak oscillators strongly depends on the applied forcing and/or coupling to neighbouring oscillators.
As we will see, the changes in the oscillation amplitudes can cause shrinkage or broadening of the entrainment range.

Figure~\ref{fig::fig01} shows the entrainment ranges of amplitude-phase models described by Eq.~(\ref{eq::complex}) driven by a periodic force of a period $T$:
\begin{equation}
\label{eq::forced}
\begin{split}
\dot z = \of{\ii\omega - \eigengamma f\of{r}}z + b\ee^{\ii\Omega t}
\end{split}
\end{equation}
The parameter $b$ denotes the {\it Zeitgeber} strength and $T = 2\pi/\Omega$ denotes the {\it Zeitgeber} period.
Depending on the {\it Zeitgeber} strength $b$, the oscillator can synchronize to a range of {\it Zeitgeber} periods $T$, also known as range of entrainment.
Entrainment range becomes broader with increasing $b$.
Thus, we find the characteristic triangular shape of the entrainment region, also termed 1:1 Arnold tongue~\cite{Alexander2009}.
For all three oscillator types, the entrainment range of the weak oscillator is broader than that of the rigid one, compare red lines against black ones in Figure~\ref{fig::fig01}.

We have identified the following mechanism that leads to the differences in the width of the entrainment range:
Generically, the entrainment range of limit-cycle oscillators is bracketed by a pair of either saddle-node (SN) or Neimark-Sacker (NS) bifurcation lines.
In rigid oscillators, SN bifurcation lines (black lines in Figure~\ref{fig::fig01}) continue up to relatively large values of $b \sim 0.05$, where the entrainment range is about 10 hours wide.
Contrarily, in weak oscillators, SN bifurcation is found only for small values of $b$ (parts of red lines {\it below} red dots in Figure~\ref{fig::fig01}).
With $b$ that small, SN lines of week and rigid oscillators coincide, compare Figure~\ref{fig::fig01} (b) and (c) with black lines occluding red ones below red dots.
For larger values of $b$, the entrainment range of weak oscillators is limited by NS lines (parts of red lines {\it above} red dots in Figure~\ref{fig::fig01}), which span a broader range in {\it Zeitgeber} period $T$.
Thus, we attribute the broader entrainment range in weak oscillators to switching from SN bifurcation to NS bifurcation, which in our case is achieved for small $\gamma$ (rigid vs. weak oscillators).
Note that in general, the switching between SN to NS bifurcation can be controlled by different parameters.

Saddle-node and Neimark-Sacker bifurcation lines can cross in codimension-two Bogdanov-Takens points \--- those are exactly the red dots in Figure~\ref{fig::fig01}.
An in-depth discussion of such high-codimension bifurcation points falls beyond the scope of the present paper and we refer to a standard textbook~\cite{Alexander2009}.


\begin{figure}
\includegraphics[width=\linewidth]{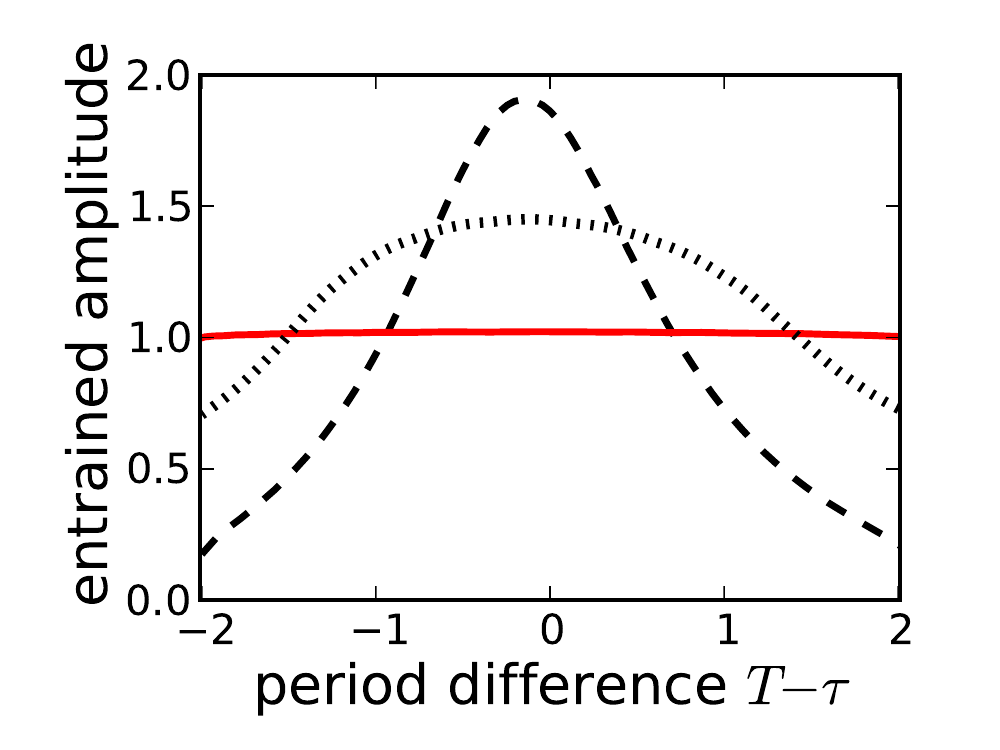}
\caption{
Dependence of the amplitude of the entrained oscillator on the detuning $T-\tau$.
The solid red line refers to a rigid linear oscillator ($\eigengamma = 1$), the dashed black line represents a weak linear oscillator ($\eigengamma = 10^{-2}$), and the dotted black line refers to weak Hopf oscillator ($\eigengamma = 10^{-2}$).
In each case, the forcing strength $b$ is chosen in such a way that the width of the synchronization range is $4$ h:
for weak Hopf oscillator $b=1.6\times 10^{-2}$, for rigid linear $b = 2.2\times 10^{-2}$ and for weak linear $b=9\times 10^{-3}$.
The Poincar\'e oscillator shows comparable behaviour.
}
\label{fig::fig02}
\end{figure}

The amplitude of entrained oscillators within the entrainment range is exemplarily shown in Figure~\ref{fig::fig02}.
As we can see, weak oscillators demonstrate a pronounced amplitude increase close to the middle of the entrainment range.
Close to the limits of entrainment, the amplitude of weak oscillators drops below the value of the amplitude of the unperturbed oscillator \cite{Wever1972}.
Amplitude changes in rigid oscillators do not show reasonable changes over the whole entrainment range.

As intuitively expected, the ratio of {\it Zeitgeber} strength $b$ to oscillator amplitude $A_0$ determines the entrainment range.
This has been confirmed in \cite{Abraham2010} using extensive simulations and analytical calculations. 
Figures~\ref{fig::fig01} and~\ref{fig::fig02} suggest that the change in the entrainment range can be also influenced by the entrained amplitude and the rigidity.
Thus we conclude that for a given entrainment strength, all three oscillator properties (the intrinsic period $\tau$, amplitude $A_0$, and the relaxation rate $\eigengamma$) shape the entrainment range.

In the following sections we take into account that the SCN oscillator is in fact a network of coupled single oscillators.
Consequently, we address the question how coupling influences the amplitude and the Floquet exponents of the synchronized SCN.
We show that mean-field coupling reduces the entrainment range via amplitude expansion.
Moreover, for identical oscillators diffusive coupling affects rigidity but causes little effects on the amplitude and therefore no major changes in the entrainment range.

\subsection{Mean-field coupling leads to amplitude expansion}
\label{Subsection::MeanField}
So far, we have considered the SCN as a single limit cycle oscillator.
In the following we take into account that the SCN is actually a network of coupled cells. 
Under normal condition these cells are well synchronized. Specific conditions such as constant light \cite{Schwartz2009} or exotic short light-dark cycles might lead to desynchronization \cite{Granada2010}.
Here, we focus on synchronized cells driven by external rhythms.
As discussed above, multiple coupling mechanisms contribute to synchronization.
We study here diffusive coupling, which represents, e.g., gap junctions, and mean-field coupling, which might model secreted neuropeptides such as VIP \cite{Welsh2010}.

\begin{figure}
\includegraphics[width=\linewidth]{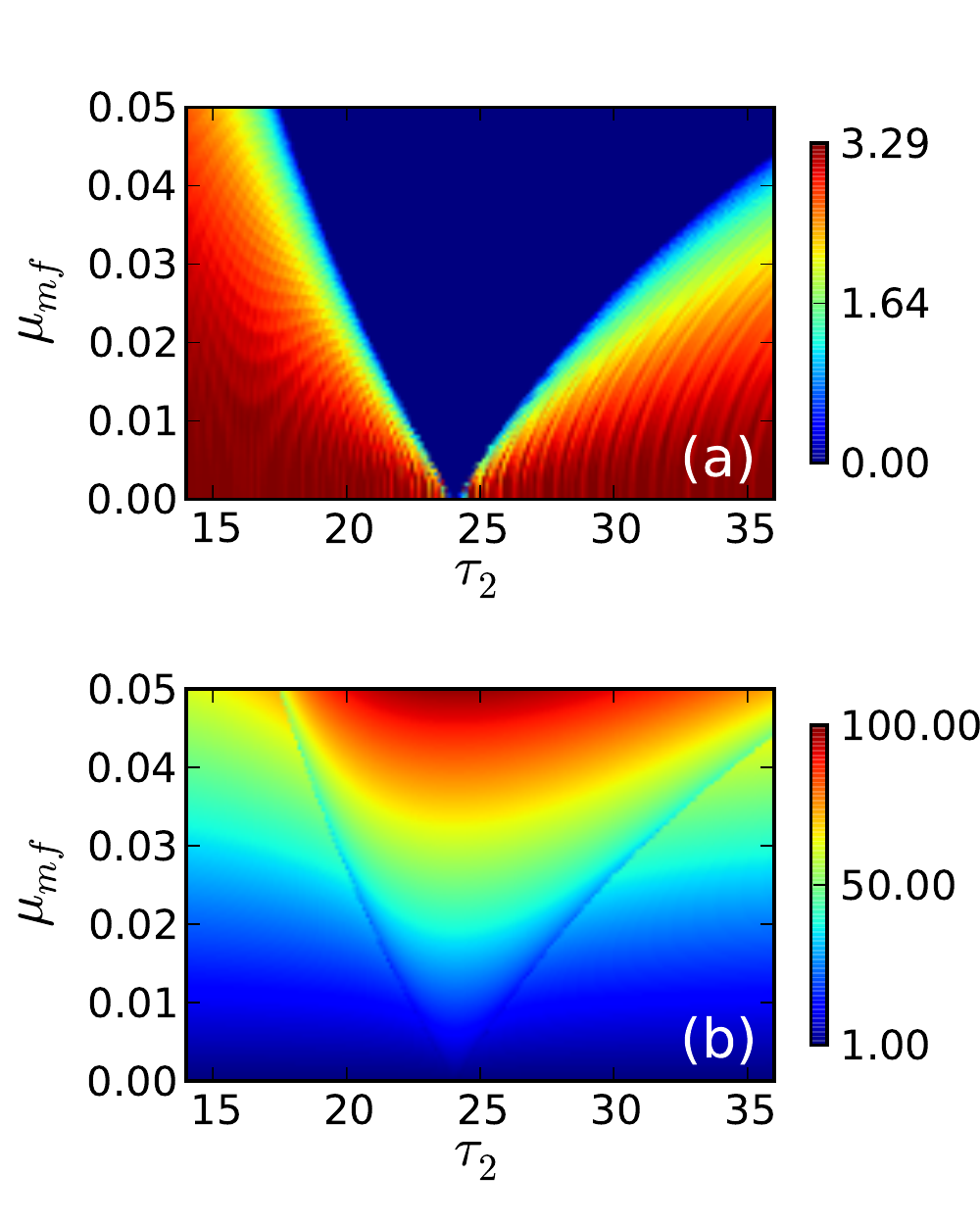}
\caption{
(a) Color-coded standard deviation of phase difference between two mean-field coupled Poincar\'e oscillators ($\gamma = 10^{-3}, A_0 = 1.0, \tau_1 = 24$).
(b) Color-coded amplitude of two mean-field coupled Poincar\'e oscillators.
Amplitudes of both oscillator are identical.
      }
\label{fig::fig03}
\end{figure}

In case of two coupled cells these coupling mechanisms are described by the following equations for complex amplitudes $z_{1,2} = r_{1,2}\ee^{\ii\varphi_{1,2}}$ of both oscillators
\begin{equation}
\label{eq::mf}
\begin{split}
\dot z_1 &= \of{\ii\omega_1 - \eigengamma f\of{r_1}} z_1 + \mu_\text{mf}\of{z_1 + z_2} + \mu_\text{d}\of{z_2 - z_1},\\
\dot z_2 &= \of{\ii\omega_2 - \eigengamma f\of{r_2}} z_2 + \mu_\text{mf}\of{z_1 + z_2} + \mu_\text{d}\of{z_1 - z_2}.
\end{split}
\end{equation}
Here, $\omega_{1,2} = \frac{2\pi}{\tau_{1,2}}$ are the oscillator frequencies with $\tau_{1,2}$ close to 24 h being their internal periods, $\mu_\text{mf}$ accounts for the strength of mean-field coupling, and $\mu_\text{d}$ accounts for the strength of diffusive coupling.
In this formulation, parameters $\mu_\text{mf}$ and $\mu_\text{d}$ can be varied independently.
Pure diffusive coupling corresponds to $\mu_\text{mf} = 0$ and pure mean-field coupling to $\mu_\text{d} = 0$.

We have discussed in the previous section that the ratio of external forcing to oscillator amplitude determines the entrainment range.
Thus for a given {\it Zeitgeber} strength $b$, the amplitudes of the coupled oscillators are essential.
Large amplitudes oscillator are difficult to entrain \cite{Pittendrigh1991,Brown2008}.

In identical oscillators that are completely synchronized (i.e. if $z_1 = z_2$), the diffusive coupling term $\mu_\text{d}\of{z_1 - z_2}$ in Eq.~(\ref{eq::mf}) vanishes and, hence, the synchronized state is the same as in an uncoupled system.
In other situations such as detuning of the frequencies, amplitude reduction can be achieved \cite{BarEli1984,aronson1990amplitude}.

Contrarily, pure mean-field coupling (i.e. $\mu_\text{d} = 0$) can induce pronounced resonance behaviour.
The coupling term $\mu_\text{mf}\of{z_1 + z_2}$ in Eq.~(\ref{eq::mf}) acts like a periodic driving force and can enhance in this way oscillation amplitudes of $z_1$ and $z_2$ drastically.
This effect is particularly strong if the oscillators are weak (small $\eigengamma$) and especially for Poincar\'e oscillator and oscillator with linearized dynamics, compare section~\ref{Subsection::Range}.

We derive in the Appendix A an analytical expression for the amplitude of the synchronized system.
It turns out that mean-field coupling generally expands amplitudes, particularly in the middle of the Arnold tongue.
Contrarily, diffusive coupling reduces the amplitude and might even suspend the oscillations completely.
Both effects are pronounced for weak oscillators, i.e. for small values of $\gamma$.

Such an amplitude expansion under mean-field coupling is illustrated in Figure~\ref{fig::fig03} (b).
The 1:1 Arnold tongue is clearly visible in the plot of the phase difference in Figure~\ref{fig::fig03}~(a).
Figure~\ref{fig::fig03}~(b) shows an amplitude expansion by two orders of magnitude, particularly within the Arnold tongue.
In section~\ref{Subsection::CouplingGovernsEntrainmentRange} we will exemplify how such an amplitude expansion leads to a drastic shrinkage of the entrainment range.

In the following section we will demonstrate that coupling strength is also intimately related to the rigidity of the coupled system as a whole.
Consequently, coupling might control the entrainment range in two ways: Via amplitude (the results of this section) and via rigidity as discussed below.

\subsection{Coupling strength determines rigidity of the coupled system}
\label{Subsection::CouplingStrengthRigidity}
We have seen in section~\ref{Subsection::Range} that beside period and amplitude, relaxation rates (the rigidity of oscillators) affect the entrainment range.
Below we will demonstrate that in a network of coupled oscillators the slowest relaxation rates (the Floquet exponents of the synchronized system as a whole) are governed by the coupling strength.

Let us consider formally an ensemble of $N$ identical uncoupled oscillators.
This large system obviously possesses $N$ zero Floquet exponents which are just the trivial Floquet exponents of the member oscillators.
Now, by introducing a small coupling, we expect that the $N$ zero Floquet exponents of the huge system slightly move away from zero, while remaining in a small vicinity of zero (except for one zero exponent which remains at zero due to the phase shift invariance).
For small coupling, the shift of Floquet exponents from zero will be proportional to the coupling strength.
Thus, the slowest time scale in the system will be dictated by the interaction between oscillators.

In Appendix B we present the matrix $M$ whose eigenvalues approximate the Floquet exponents of the synchronized state of weakly coupled oscillators.
This matrix has the following properties:
It applies to a general setup of an arbitrary number of limit-cycle oscillators of any dimension and is universally applicable to different models of circadian clocks, including the simple models considered in this paper.
Matrix $M$ can be deduced just from the properties of the unperturbed limit cycle oscillator.
This in turn implies that using matrix $M$, it is possible to calculate the rigidity of the coupled system as a whole only from information on the single uncoupled oscillator.
This matrix depends only on how the coupled oscillators cross-talk to each other.
In this sense, this matrix is applicable for both mean-field and diffusive coupling.

As a main result, we find that the relaxation rates (the Floquet exponents) of the synchronized system are proportional to the coupling strength for almost any oscillator model and coupling type.

\begin{figure}
\includegraphics[width=\linewidth]{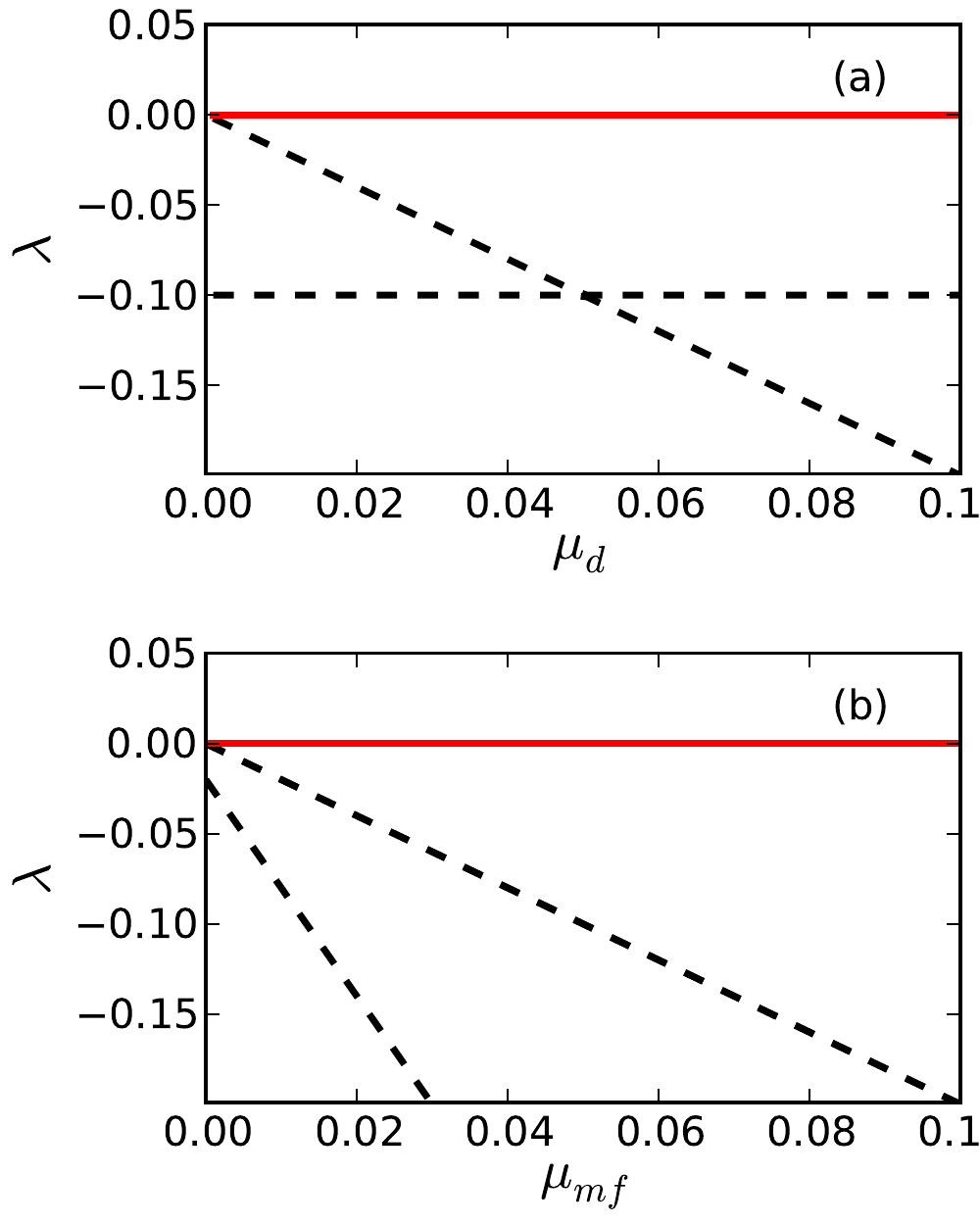}
\caption{
Dependence of the Floquet exponents $\lambda$ on coupling strength.
      (a) Two diffusively coupled (strenth $\mu_\text{d}$) Poincar\'e oscillators with $\eigengamma = 10^{-1}$.
        Dashed horizontal line denotes the local Floquet exponent which is not affected by the coupling strength.
          In contrast, the other sloped dashed line denote the interaction Floquet exponent that decreases with increasing coupling strength.
      (b) Two mean-field coupled (strength $\mu_\text{mf}$) Hopf oscillators with $\eigengamma = 10^{-2}$.
      In both (a) and (b), red straight lines denote the trivial Floquet exponent at zero.
      Due to the amplitude expansions, both non-trivial Floquet multipliers (dashed lines) decrease as coupling strength increases.
        Also, there always exists a fourth exponent at much smaller values of $\lambda$, which is not included in (a) and (b).
}
\label{fig::fig04}
\end{figure}

Figure~\ref{fig::fig04} illustrates the linear decay of Floquet exponents with coupling strength.
For diffusive coupling as in Figure~\ref{fig::fig04}~(a), homogeneous perturbations decay with the rate $\gamma$ (the horizontal line in Figure~\ref{fig::fig04}~(a)), since the coupling terms $z_{1,2} - z_{2,1}$ vanish due to the symmetry.
As intuitively expected, heterogeneous perturbations decay with a rate proportional to the coupling strength.
Consequently, monitoring of transients in experiments can potentially provide information on coupling strength.
Figure~\ref{fig::fig04} (b) shows the Floquet exponents of the mean-field coupled system.
Beside the trivial exponent at zero, there are two small negative exponents that grow into negative values as the coupling strength $\mu_\text{mf}$ increases.
In contrast to Figure~\ref{fig::fig04} (a), the single oscillator relaxation rate $\gamma$ does not persist at a constant value as coupling strength increases.
We explain this growth of the absolute value of the exponents by the fact that increasing mean-field coupling leads to an increase of the amplitude of the synchronized state (see for instance Figure~\ref{fig::fig03} (b)).
Due to the nonlinearity of the system, the exponent depends on the size of the limit cycle and thus the increasing limit cycle becomes more stable.
This is of course not the case for diffusive coupling, since diffusive coupling does not lead to an amplitude increase (see Figure~\ref{fig::fig04} (a)).

\subsection{Coupling governs entrainment range}
\label{Subsection::CouplingGovernsEntrainmentRange}
We have shown above that coupling strength determines amplitudes and Floquet exponents of a coupled oscillator system.
It has been argued in Introduction that entrainment to external rhythms is crucial for an organism's fitness.
It was found experimentally that coupled SCN neurons are harder to entrain than peripherial tissues \cite{Buhr2010,Abraham2010}.
We illustrate in this section that rigidity and amplitude expansion affect the widths of the Arnold tongues.

Figure~\ref{fig::fig05} shows the entrainment ranges of coupled rigid (a, b) and weak (c,d) oscillators.
As a model, we took coupled Eqs.~(\ref{eq::mf}) with right-hand sides for $z_1$ and $z_2$ perturbed by the same periodic forcing in the form of $b\ee^{\ii \frac{2\pi}{T} t}$ with the amplitude $b$ and period $T$.
It turns out that coupled rigid cells in Figure~\ref{fig::fig05} (b) show no amplitude expansion and, consequently, the Arnold tongue resembles the entrainment zones in Figure~\ref{fig::fig01}.
In contrast, amplitude expansion of weak oscillators leads to a drastic shrinkage of the Arnold tongue, see Figure~\ref{fig::fig05}(c) and (d).
Similar effects of mean-field coupling are observed in the other oscillator models (the Hopf oscillator and the oscillator with linear radial dynamics).

\section{Discussion}
\label{Section::Discussion}
Circadian oscillators have strikingly similar molecular mechanisms both in SCN neurons and in peripheral cells \cite{Liu2007a,Asher2011}.
At single cell level, the relative amplitudes vary from $0.1$ to $2$ and estimated damping rates are in the range from $0.001$ h$^{-1}$ to $0.3$ h$^{-1}$ \cite{Westermark2009}.
Single cell periods have a standard deviation of 1-2 h \cite{Welsh1995,Herzog2004}.
Despite such variable and noisy single cell oscillations, the SCN as a whole is a really precise pacemaker with a day-to-day variation of a few minutes \cite{Herzog2004}.
This precision is believed to be achieved by intercellular coupling of SCN neurons \cite{Welsh2010}.

Recent experiments revealed that coupling governs the entrainment range of circadian clocks \cite{Buhr2010,Abraham2010}.
Lung tissue without strong intercellular coupling was entrained by 1.5$^{\circ}\text{C}$ temperature cycles at external periods of 20 h and 28 h.
Contrarily, SCN tissue could not be entrained even by a larger temperature amplitude of $4^{\circ}\text{C}$ and an external period of 22 h \cite{Abraham2010}.
Only if the coupling was reduced via chemicals (MDL and TTX), entrainment was achieved.
These experimental results motivated our systematic analysis of coupling and entrainment.

We have shown in this paper that coupling can affect the oscillator properties of the network in two ways: {\em (i)} via amplitude expansion due to mean-field coupling and {\em (ii)} via increased rigidity of the SCN network.
Both effects have been suggested to explain the differences between lung and SCN tissues \cite{Abraham2010}.

\begin{figure*}
\includegraphics[width=\linewidth]{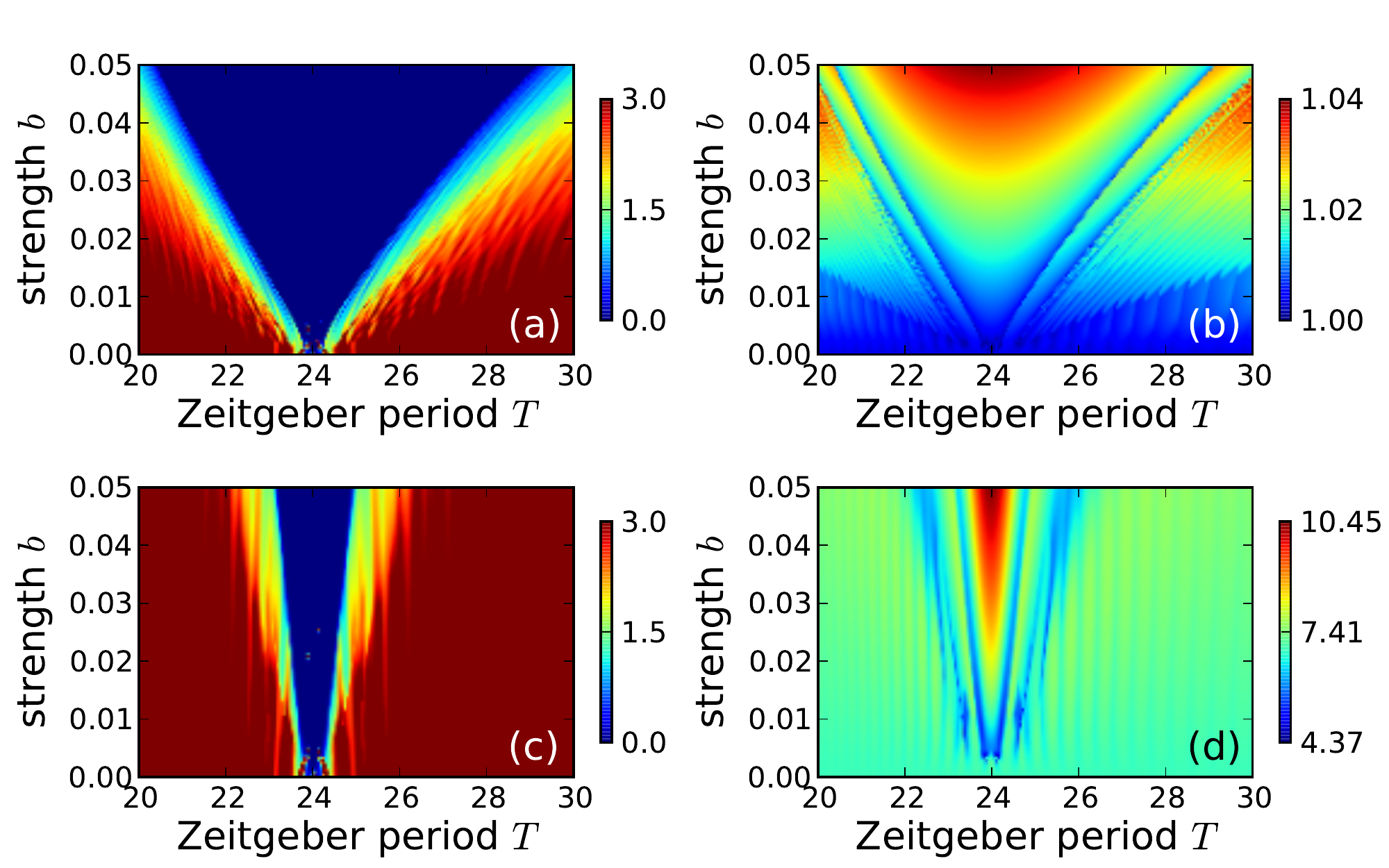}
\caption{
Phase difference and amplitude behaviour of two mean-field coupled Poincar\'e oscillators under external forcing.
Color-coded standard deviation of the phase difference (left panels) and entrained amplitude (right panels).
Both coupled oscillators are subjected to the same external {\it Zeitgeber} period $T$ and {\it Zeitgeber} strength $b$.
Further parameters are: $A_{0_{1,2}} = 1, \mu_\text{mf} = 0.005, \tau_1 = 23.5, \tau_2 = 24.5$,
(a) and (b) rigid oscillators ($\gamma_{1,2} = 1$).
Note the W-shaped Arnold tongue due to the fact that $\tau_1 \neq \tau_2$ and hence the {\it individual} entrainment ranges of both oscillators are shifted with respect to each other.
(c) and (d) weak oscillators ($\gamma_{1,2} = 10^{-3}$).
      }
\label{fig::fig05}
\end{figure*}

Our results on the Floquet multipliers of the synchronized state in ensembles of oscillators have been obtained in a quite general setup for an arbitrary number of oscillators and connection topology.
We also strongly believe that our numerical results can be straight-forwardly generalized to a larger number of interacting oscillators.
\\
On the other hand, the structure of Arnold tongue becomes more involved for oscillator ensembles with different internal frequencies.
Figure~\ref{fig::fig05} suggest that already two oscillators result in a W-shaped entrainment range.
We also speculate that the visible rippling in Figure~\ref{fig::fig05} might be attributed to a secondary bifurcation structure due to the mismatch of the internal periods.

For a quantitative comparison of our simulations with the in vivo SCN network, direct measurements of the coupling strength and oscillator rigidity are desirable.
However, such data are currently not available.
There are chemical treatments with TTX \cite{Yamaguchi2003} and MDL \cite{ONeill2008}, studies with dissociated neurons \cite{Herzog2004,Liu2007a} and mechanically cut slices \cite{Yamaguchi2003}.

Furthermore, knockout studies have been performed extensively (VIP, VIPR, gap junctions).
Unfortunately, all these interventions lead to considerable destruction of the SCN network and to poorly controlled side effects such as changes of ion and neurotransmitter concentrations.

Thus coupling mechanisms in the intact SCN have to be explored indirectly.
Using powerful imaging techniques, amplitudes and phases of single cells can be monitored~\cite{Liu2007a,Abraham2010}.
Temperature cycles allow the analysis of entrainment properties.
Moreover, temperature pulse response can be used to derive phase response curves \cite{Buhr2010,Granada2009a}.
Such indirect measurements can be exploited to infer properties of single cell oscillators and their coupling.
This reverse approach can profit from the theory presented in this paper.
There are, for instance, indications that coupled SCN neurons exhibit larger amplitudes than dissociated neurons \cite{Liu2007a,Westermark2009}.
This could reflect amplitude expansion via mean-field coupling as studied in section~\ref{Subsection::MeanField}.
The enlarged  entrainment range due to MDL treatment in \cite{Abraham2010} might be related to reduced rigidity, since the relative amplitudes are unaffected.

In summary, our results suggest that the established theory of coupled oscillators can provide valuable insights in the field of circadian rhythms.

\appendix
\section{Entrained amplitude for mean-field and diffusive coupling}
The aim of this appendix is to demonstrate how mean-field and diffusional coupling together with the amount of detuning affects the amplitude of the synchronized solution.
Our calculations are based on the method described in \cite{aronson1990amplitude}.

Rewriting Eq.~(\ref{eq::mf}) in polar form with $z_{1,2} = r_{1,2}\ee^{\ii\varphi_{1,2}}$ and looking for a stationary solution with $r_1 = r_2 = r$, we arrive at
\begin{equation}
\label{eq::polarSym}
\begin{split}
\dot r & = \of{\mu_m + \mu_p\cos{\theta} - \eigengamma f\of{r}}r,\\
\dot \theta &= \Delta - 2\mu_p \sin\theta,
\end{split}
\end{equation}
where we have defined $\theta = \varphi_1 - \varphi_2$, $\Delta = \omega_1 - \omega_2,$
$\mu_p = \mu_\text{mf} + \mu_\text{d}$, and
$\mu_m = \mu_\text{mf} - \mu_\text{d}$.

Looking for phase-locking regimes is equivalent to $\dot \theta = 0$, which immediately results in $\sin\theta = \frac{\Delta}{2\mu_p}$.
We can use this to express $\cos\theta$ in the first equation of Eq.~(\ref{eq::polarSym}) and impose $\dot r = 0$, which results in either $r=0$ or
\begin{equation}
\label{eq::symmetric}
f\of{r} = \of{\mu_m \pm \sqrt{\mu_p^2 - \frac{\Delta}{4}}}/\eigengamma
\end{equation}
Here, $f\of{r}$ is one of the three possible radial dynamics from Eq.~(\ref{eq::threechoices}).
Thus, Eq.~(\ref{eq::symmetric}) constitutes the conditional equation for the amplitude $r$ of the synchronized state.
We see that both coefficients for mean-field and diffusive coupling contribute to the amplitude change.
Note that one of the solution (with plus sign) is a stable one, whereas the second one is unstable.

Some special cases are worth mentioning here.
Consider first two identical oscillators with $\Delta = 0$, then the above expression for the amplitude $r$ simplifies to
\[
f\of{r} = \of{\mu_m \pm \mu_p}/\eigengamma
\]
which results in either $f\of{r} = 2\mu_\text{mf}/\eigengamma$ or $f\of{r} = 2\mu_\text{d}/\eigengamma$.
This explains the amplitude expansion for the case of pure mean-field coupling with $\mu_\text{d} = 0$.
Moreover, choosing $\mu_{m,p}$ such that $\of{\mu_m + \mu_p}/\eigengamma \leq -1$ ``kills'' the linearly unstable part of $f\of{r}$ and thus results in an absence of oscillations~\cite{BarEli1984}.

The condition for a phase-locking solution $\sin\theta = \frac{\Delta}{2\mu_p}$ describes the width of the Arnold tongue, giving $-2\mu_p \leq \Delta \leq 2\mu_p$.
At the border of the entrainment range, both stable and unstable solutions disappear in a saddle-node bifurcation.
For given $\mu_p$ and $\mu_m$, the amplitude of the stable solution (with plus sign) has a maximum at $\Delta=0$, which follows from the inspection of Eq.~(\ref{eq::symmetric}).

\section{Floquet exponents of fully synchronized state of identical oscillators}
Let us consider $N$ identical oscillators, each of them described by a $\ell$-dimensional state vector $u_i \in \mathbb{R}^\ell$, $i = 1,2, \dots, N$.
Suppose that in the absence of coupling each of the oscillators is described by the equation
\begin{equation}
\label{eq::single}
\begin{split}
\dot u = g\of{u}.
\end{split}
\end{equation}
The unperturbed limit cycle is $u\of{t} = q\of{t}$ with period $\tau$ such that $q\of{t + \tau} = q\of{t}$ for all $t$.
Now we shall study a coupled ensemble of $N$ identical oscillators
\begin{equation}
\label{eq::coupled}
\begin{split}
\dot u_i = g\of{u_i} + k_i \of{u_1, u_2, \dots, u_N; \epsilon}, \quad i = 1, 2, \dots, N,
\end{split}
\end{equation}
where $k_i\of{u_1, u_2, \dots, u_N; \epsilon}$ are coupling functions depending on the coupling strength $\epsilon$. For small $\epsilon$, the coupling functions $k_i$ are also assumed to be small, i.e. $k_i\of{\dots; \epsilon} = \mathcal{O}\of{\epsilon}$.
A straight-forward, but rather tedious perturbation calculation in small parameter $\epsilon$ shows that the $N\times\ell$-dimensional limit cycle oscillator exhibits $N-1$ Floquet eigenvalues being $\epsilon$-small.
One Floquet eigenvalue is zero, as implied by the phase shift invariance.
The $N$ critical (those close to zero) Floquet exponents of Eq.~(\ref{eq::coupled}) are approximated to the $\epsilon$ order by the eigenvalues of the following matrix $M$:
\begin{equation}
\label{eq::MatrixW}
\begin{split}
M = \begin{pmatrix}
{\displaystyle -\sum_i^{i \neq 1} \kappa_{1i}} & \kappa_{12} & \dots & \kappa_{1N} \\
\kappa_{21} & {\displaystyle -\sum_i^{i \neq 2} \kappa_{2i}} & \dots & \kappa_{2N}\\
\vdots & \vdots & \ddots & \vdots\\
\kappa_{N1} & \kappa_{N2} & \dots & {\displaystyle -\sum_i^{i \neq N} \kappa_{Ni}}
\end{pmatrix}.
\end{split}
\end{equation}
The coefficients in the matrix depend on the integrals
\[
\kappa_{ij} = \int\limits_0^\tau \dd t \, \psi\of{t} K_{ij}\of{t;\epsilon} \dot q\of{t},
\]
where $\psi\of{t}$ is the eigenfunction to the zero eigenvalue of the adjoint linearization of Eq.~(\ref{eq::single}) around the unperturbed limit cycle $q\of{t}$ and $K_{ij} = \fracp{k_i}{u_j}$ are the Jacobian matrix elements of the coupling functions $k_i$ evaluated along $q\of{t}$.
Note that the matrix $M$ does not contain terms $\kappa_{ii}$. Those reflect the influence of the $u_i$ on itself through the coupling function $k_i$.
Thus, for both mean-field coupling and diffusive coupling the matrix $M$ will be the same.

If all coupling functions $k_i\of{u_1, u_2, \dots, u_N; \epsilon}$ depend on all other $u_j, j \neq i$ in the same manner, that is, all $\fracp{k_i}{u_j}$ are equal, the matrix $M$ simplifies to
\[
M = 
\kappa_o
\begin{pmatrix}
-\alpha & \beta   & \dots  & \beta\\
\beta   & -\alpha & \dots  & \beta\\
\vdots  & \vdots  & \ddots & \vdots\\
\beta   & \beta   & \dots  & -\alpha
\end{pmatrix}
\]
with 
$$
\kappa_o = \int\limits_0^\tau \dd t \, \psi\of{t} \fracp{k_i}{u_j}\of{t;\epsilon} \dot q\of{t},
$$
and $\alpha = (N - 1)/N$ and $\beta = 1/N$.
For $k_i\of{u_1, \dots, u_N; \epsilon} = \epsilon N^{-1} \sum_j^N M_0 u_j$ we have $\fracp{k_i}{u_j} = \epsilon N^{-1} M_o$ for all $i\neq j$, where $M_o$ is a $\ell \times \ell$ matrix with constant coefficients that describe how each of $\ell$ components of $u$ are coupled to each other.
Hence, $\kappa_o$ is just a scalar proportional to $\epsilon$.

The eigenvalues of $M$ are given by zero and $N-1$ eigenvalues at $-\kappa_o$.
The eigenvector to the zero eigenvalue corresponds to a simultaneous shift of all $N$ phases by the same amount.
The eigenvectors to the non-zero eigenvalues are given by the periodic shifts of the vector $\of{N-1, -1, -1, \dots, -1}^+$.
In the limit of large $N$ those eigenvectors correspond to the ``evaporation'' eigenvalue: they describe the stability of the synchronized state against perturbations of a single oscillator, see~\cite{PopovychPRL2001}.

\section{Numerical Methods}
\label{Section::Methods}
Direct numerical simulations were performed with the help of the explicit 4th-order Runge-Kutta method with a time step of $\Delta t = 0.01$ d.u. (compare that to the typical period of oscillation $\tau = 24$ d.u.).
Averaging for each parameter point in Figures~\ref{fig::fig03} and \ref{fig::fig05} was done over the last 40 periods of total 140 periods of simulated oscillations.
Bifurcation lines in Figures \ref{fig::fig01}, \ref{fig::fig02}, and Floquet exponents in Figure \ref{fig::fig04} were obtained with the help of the standard numerical continuation and bifurcation analysis software AUTO 2000 \cite{AUTO}.

\bibliographystyle{epj}
\bibliography{coupled}

\end{document}